\documentclass[aps,floatfix,twocolumn,showpacs,preprintnumbers]{revtex4}
\usepackage{graphicx}
\usepackage{bm}
\usepackage{amsmath,amssymb}
\usepackage{xcolor}

\graphicspath{ {./}{figs/}}
\begin{document}

\title{Dirac Semimetal Phase in Rhombohedral $\beta -$Cu$_{2}$Se}
\author{Thomas Steele$^{\dagger }$, Becker Sharif$^{\ddagger }$, David
Lederman$^{\ddagger }$, Xiangang Wan$^{\ast }$ and Sergey Y. Savrasov$^{\ast
,\dagger }$}
\affiliation{$^{\dagger }$Department of Physics, University of California, Davis, CA
95616, USA}
\affiliation{$^{\ddagger }$Department of Physics, University of California, Santa Cruz,
Santa Cruz, CA 95064, USA and Materials Science and Engineering Program,
University of California, Santa Cruz, Santa Cruz, CA 95064, USA}
\affiliation{$^{\ast }$National Laboratory of Solid State Microstructures, School of
Physics and Collaborative Innovation Center of Advanced Microstructures,
Nanjing University, Nanjing 210093, China}

\begin{abstract}
Having been extensively studied during last decades in the fields of
thermoelectics and ionic conductors, the $\alpha $ phase of Cu$_{2}$Se with
antfluoride crystal structure has recently emerged as a topological
zero--gap semimetal with a quadratic contact point which exists at the Fermi
surface of its bulk electronic spectrum. Here we argue based on density
functional electronic structure calculation that the $\beta $ phase of Cu$%
_{2}$ Se realized in a recently discovered rhombohedral structure shows a
Dirac semimetal behavior of the electrons near the Fermi level. These
topological semimetals are currently generating a lot of interest due to
unusual transport phenomena, such as strong quantum oscillations, large
magnetoresistance effect and ultrahigh carrier mobilities with their Fermi
velocities potentially exceeding graphene. We show that there exist Fermi
arc states at the surface spectrum of $\beta -$Cu$_{2}$Se that are
topologically protected by the bulk Dirac points. Their shape and spin
properties should be resilient to the back- and side scattering effects in
the surface transport, suggesting new ways for realizing high-mobility
electronic devices.
\end{abstract}

\maketitle

\section{\textbf{Introduction.}}

Copper selenide, a material widely regarded as a possible substitute for Li
batteries, exhibits a remarkable liquid--like conductivity of the Cu ions 
\cite{Yakshibaev,Korzhuev} and at the same time shows a high thermoelectric
effect \cite{Snyder}. It most often occurs in nature with cation deficiency
described by chemical formula Cu$_{2-\delta }$Se where the range of $\delta $
is between $0$ and 0.25. There are two known crystallographic phases, $%
\alpha $ and $\beta ,$ and the transition between them occuring around 414K 
\cite{Abrikosov}, has recently attracted a lot of research interest\cite%
{A1,A2,A3,A4,A5,Zhao}. Although at temperatures above the transition the
antifluoride structure of $\alpha -$Cu$_{2}$Se with the space group $Fm3m$
is well established\cite{Skomorokhov}, the structure of $\beta $--Cu$_{2}$Se
at low temperatures has been controversial in part due to
non--stoichiometricity and random distribution of Cu ions. It was proposed
in the past to be either monoclinic, orthorhombic, or tetragonal \cite%
{Milat,Kashida,Frangis}. Recent single crystal x--ray diffraction
experiments, however, revealed the average crystal structure of the $\beta $
phase to be rhombohedral with space--group symmetry $R\bar{3}m$\cite{SCXRD}.

Using density functional theory (DFT)\cite{DFT} based electronic structure
calculation, the stoichiometric $\alpha -$Cu$_{2}$Se was shown to be a
zero--gap semiconductor \cite{Delin}. This conclusion however contradicted
earlier experimental studies that have observed an optical gap of 1.23 eV 
\cite{Sorokin}. A variety of approximations to exchange--correlation effects
has been explored to understand the lack of gap opening in the theoretical
calculation that included popular local density approximation (LDA)\cite{DFT}%
, the generalized gradient approximation (GGA)\cite{GGA}, the so--called
AM05 functional \cite{AM05} as well as hybrid functionals \cite%
{Burke,Scuseria}.

Recent discoveries of topological quantum materials, such as topological
insulators\cite{TIReview} as well as Weyl and Dirac semimetals\cite%
{WSMReview} have brought a new insight to the electronic structure of the
antifluoride $\alpha -$Cu$_{2}$Se. It turns out that the zero gap calculated
earlier in Ref. \cite{Delin} is due to the presence of a contact point at $%
\mathbf{k}=0$ between valence and conduction energy bands at the Fermi level
whose dispersion is quadratic in all three directions\cite{Weng}. It has
been shown that depending on the inclusion of spin--orbit coupling, triply
or quadruply degenerate quadratic contact points (QCPs) that are protected
by the crystalline symmetry can be realized in materials that\ should
exhibit some unconventional features in the Landau spectrum under a strong
magnetic field\cite{Weng}. The cubic $\alpha -$Cu$_{2}$Se provides an
example of such QCP\ semimetal from where, by breaking symmetry either via
the Zeeman field or by applying a lattice strain, several topological phases
can be derived, such as a Weyl semimetal , a Z$_{2}$ topological insulator
or metal, as well as a Dirac semimetal.

Here we argue that the realization of the Dirac semimetal phase occurs
naturally in the $\beta $ phase of Cu$_{2}$Se. Based on our density
functional calculation, the Dirac dispersions persist along $k_{z}$%
--direction in the Brillouin Zone (BZ) of the rhombohedral crystal structure
with two Dirac points located in the vicinity of the $\mathbf{k}=0$ $\Gamma $
point and pinned at the Fermi level. Our calculation for the surface states
reveals the existence of the Fermi arcs \cite{Arcs} that connect the Dirac
points in the surface BZ. As we have discussed recently\cite{TaAs,Cd3As2},
the shape and spin properties of the Fermi arcs usually suppress both back--
and side scattering, the effect leading to very high mobilities of the
surface electrons. This has been already observed in transport studies of
NbAs Weyl semimetal\cite{NbAsNature} and Cd$_{3}$As$_{2}$ Dirac semimetal%
\cite{Cd3As2Nature}, therefore similar unusual properties of $\beta -$Cu$%
_{2} $Se can be expected.

Our paper is organized as follows. In Section II we describe the evolution
of the crystal structure of Cu$_{2}$Se from its cubic $\alpha $ phase to the
rhombohedral $\beta $ phase. Density functional electronic structure
calculations for the bulk states of $\beta -$Cu$_{2}$Se are given in Section
III. Section IV presents calculation of the Fermi arcs surface states.
Section V is the conclusion.

\section{\textbf{Crystal Structure of} $\protect\beta -$\textbf{Cu}$_{2}$%
\textbf{Se}.}

\begin{figure}[tbp]
\includegraphics[height=0.95\textwidth,width=0.4\textwidth]{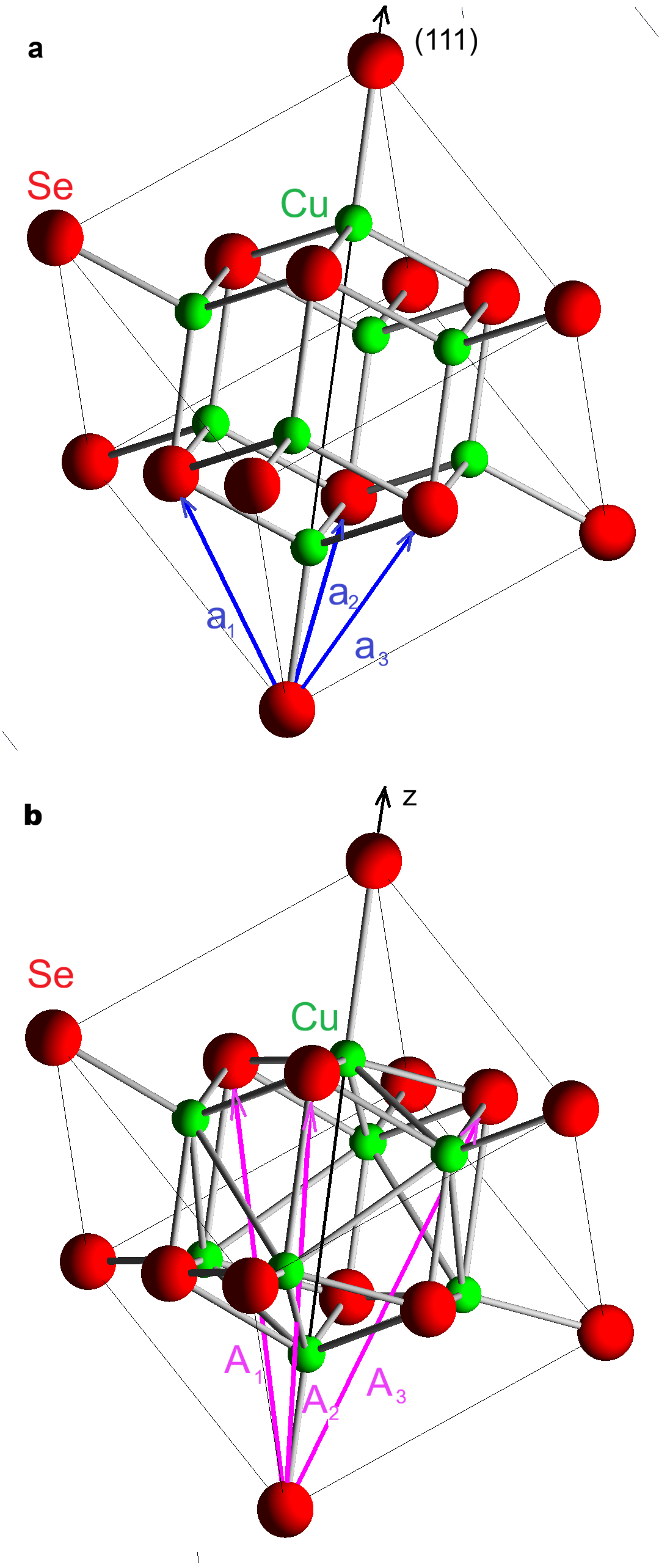}
\caption{a) Antifluoride crystal structure of $\protect\alpha -$Cu$_{2}$Se
oriented along (111) direction of the cubic lattice; The lattice
translations for the original unit cell, $\mathbf{a}_{1},\mathbf{a}_{2},%
\mathbf{a}_{3}\mathbf{,}$ are indicated. b) Rhombohedral crystal structure
of $\protect\beta -$Cu$_{2}$Se derived from the cubic phase by choosing
z--axis along $(111),$ rearranging the interlayer spacing between Se atoms
which results in doubling the unit cell of the cubic phase with new lattice
translations $\mathbf{A}_{1},\mathbf{A}_{2},\mathbf{A}_{3},$ and additional
displacements of Cu atoms from their ideal tetrahedral interstitials%
\protect\cite{SCXRD}.}
\label{FigStruc}
\end{figure}

To describe the crystal structure of the low temperature $\beta $ phase, we
first remind ourselves the antifluoride cubic structure which is realized in 
$\alpha -$Cu$_{2}$Se at high temperatures. Here Se atoms form the face
centered cubic (fcc)\ lattice while Cu atoms occupy tetrahedral
interstitials often occuring with only partial occupancy leading to the hole
doped compounds Cu$_{2-\delta }$Se. The lattice constant $a$ has been
measured to be $5.85$\AA\ \cite{SCXRD}. We have recently shown that at the
concentrations $\delta \sim 0.25$, this phase exhibits signatures of charge
density wave instability due to the Fermi surface nesting of the electronic
structure \cite{Becker}.

\begin{table}[b]
\caption{Experimentally determined crystallographic coordinates of Se and Cu
atoms in the rhombohedral $\protect\beta $ phase of Cu$_{2}$Se\protect\cite%
{SCXRD}. The primitive lattice translations $\mathbf{A}_{1},\mathbf{A}_{2},%
\mathbf{A}_{3}$ are also indicated. All coordinates are given in Cartesian
system and in the units of lattice constants $a=b=4.12$ \AA $,c=20.45$ \AA $%
. $}%
\begin{tabular}{cc}
\hline\hline
Primitive Translations & $(x,y,z)$ \\ \hline
$\mathbf{A_{1}}$ & $\frac{1}{2}$,$\frac{\sqrt{3}}{6}$,$\frac{1}{3}$ \\ 
$\mathbf{A_{2}}$ & -$\frac{1}{2}$,$\frac{\sqrt{3}}{6}$,$\frac{1}{3}$ \\ 
$\mathbf{A_{3}}$ & $0$,$-\frac{\sqrt{3}}{3}$,$\frac{1}{3}$ \\ \hline
Atomic Positions & $(x,y,z)$ \\ \hline
Se[I] & $0,0,0$ \\ 
Se[II] & $0.0,0.0,0.4815$ \\ 
Cu[I] & $0.0,0.0,0.1172$ \\ 
Cu[II] & $0.0,0.0,0.4020$ \\ 
Cu[III] & $0.0,0.0,-0.1406$ \\ 
Cu[IV] & $0.0,0.0,-0.3402$ \\ \hline\hline
\end{tabular}%
\end{table}

As we illustrate in Fig. \ref{FigStruc}(a), the fcc lattice can be viewed as
ABC stacking of the atomic Se layers along (111) direction. The transition
from cubic to rhombohedral phase occurs by slight rearrangement of the
interlayer distances which results in the new unit cell with the z--axis
pointing along (111) and also doubling the unit cell along this direction as
compared to the cubic phase, see Fig.\ref{FigStruc} (b) for illustration.
The lattice group symmetry is lowered from $Fm3m$ to $R\bar{3}m$, and the
new lattice constants obtained experimentally are as follows: $a=b=4.12$ \AA %
,$c=20.45$ \AA \cite{SCXRD}. They are indeed very close to the values for
the cubic phase should we describe the latter one in the coordinate system
with z along (111): the lattice constant $4.12$ \AA\ can be compared to its
value $5.85/\sqrt{2}=4.14$\AA\ for the cubic phase while $20.45$ \AA\ is
very close to $5.85\times 2\sqrt{3}=20.26$ \AA .

The Cu atoms occupying the tetrahedral interstitials of the fcc lattice in $%
\alpha -$Cu$_{2}$Se displace from their ideal positions which results in two
different kinds of Cu sites in the $\beta $ phase. We summarize its
experimentally determined lattice parameters in Table 1\cite{SCXRD} and
indicate old and new lattice translations in Fig. \ref{FigStruc} (a) and (b)
respectively.

To conclude this description we note that a recent comprehensive study of
the $\beta $ phase by advanced electron microscopy \cite{Zhao} has also
detected a modulation of the crystal structure in the $\beta $ phase. Let A
to refer to the unit cell of the rhombohedral phase and B to be A rotated by 
$n\pi /6$ ($n=1,3,5...$) around z--axis. These two basic units A and B form
a superstructure along $c$ by various permutations, such as ABBA, BBBA, 
\textit{etc.}. This results in a variety of random copper sublattices that
are embedded into the Se layers following the fcc stacking.

\section{\textbf{Dirac Dispersions in} $\protect\beta -$\textbf{Cu}$_{2}$%
\textbf{Se.}}

\begin{figure}[tbp]
\includegraphics[height=0.57\textwidth,width=0.4\textwidth]{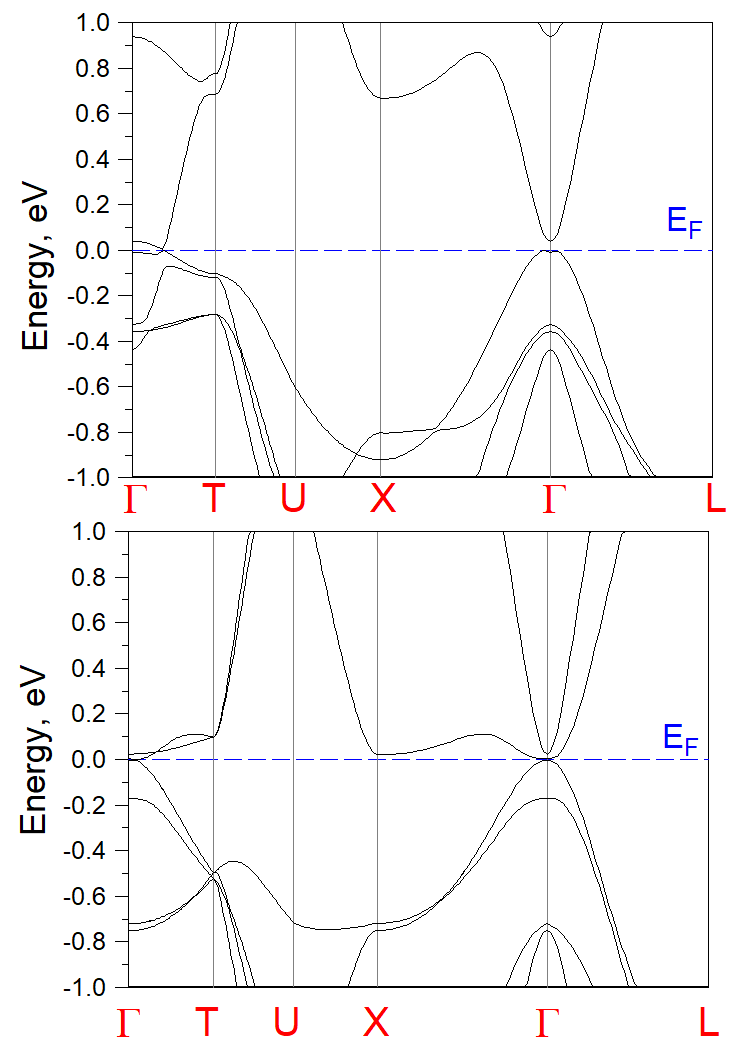}
\caption{a) Calculated band structure of the $\protect\beta $ phase of Cu$%
_{2}$Se along high--symmetry lines of the rhombohedral Brillouin Zone. b)
Energy band dispersions of the cubic $\protect\alpha -$Cu$_{2}$Se drawn in
the rhombohedral Brillouin Zone by utilizing z axis along (111) and by
doubling the unit cell to make direct comparisons between the bands of $%
\protect\alpha $ and $\protect\beta $ phases possible.}
\label{FigBands}
\end{figure}

We now turn to the description of the electronic states in $\beta -$Cu$_{2}$%
Se. We perform our density--functional electronic--structure calculations
using local density approximation and the full potential linear muffin--tin
orbital method\cite{FPLMTO} with spin--orbit coupling. The result for the
electronic energy bands is shown in Fig. \ref{FigBands}(a) along major
high--symmetry directions of the rhombohedral lattice BZ\ that is
illustrated in Fig. \ref{FigFermiArcs}(a). We emphasize the appearance of
the Dirac states along k$_{z}$ direction in the vicinity of the $\Gamma $
point which cross exactly at the Fermi level. The Dirac points are protected
by the C$_{3z}$ symmetry and their locations are indicated by blue dots in
Fig \ref{FigFermiArcs}(a). Thus, $\beta -$Cu$_{2}$Se is predicted to be
ideal Dirac semimetal in our calculation. 

We can compare the evolution of the band structure from cubic to
rhombohedral phase that results in this Dirac semimetal behavior. Density
functional calculations for the antifluoride Cu$_{2}$Se have already been
performed earlier\cite{Delin,Weng}. In particular, the existence of
quadratic contact point at $\Gamma $ was highlighted\cite{Weng} as a link to
topological quantum materials that recently received an enormous interest.
Since the cubic phase can be viewed in the rhombohedral crystalline
environment by choosing (111) direction to be its new z axis, we can make
direct comparisons of the electronic states using the same crystallographic
arrangements for both $\alpha $ and $\beta $ Cu$_{2}$Se. This is illustrated
in Fig. \ref{FigBands}(b) where the energy dispersions of $\alpha $--Cu$_{2}$%
Se are plotted in the rhombohedral BZ, Fig. \ref{FigFermiArcs}(a).

As one sees the two Kramers degenerate bands merge at $\Gamma $ and form a
contact point right at the Fermi level that disperse quadratically in its
vicinity. The ideal $c/a$ ratio equal to $2\sqrt{6}=4.899$ as derived from
the cubic lattice, changes to $c/a=4.963$ for the experimentally determined
rhombohedral structure\cite{SCXRD}. It can be interpreted as a perturbation
by a tensile strain to the cube along its main diagonal. This lifts the
degeneracy at $\Gamma $ and opens a small gap as seen from comparing Figs %
\ref{FigBands}(a) and (b). The bands are still two fold degenerate as
protected by the combined inversion and time reversal symmetry; they cross
along the $k_{z}$ direction. The crossing is protected by the presence of
three--fold rotational axis resulting in the Dirac point pinned exactly at
the Fermi level.

\section{Fermi Arcs in $\protect\beta -$Cu$_{2}$Se}

In topological materials, there exists a bulk--boundary correspondence that
assumes the occurrence of the surface states that are topologically
protected by the properties of the bulk \cite{RMPTI}. In Weyl semimetals,
these are the Weyl nodes that are characterized by the chirality, and the
non--zero Berry flux through any surface surrounding each Weyl point
guarantees the existence of the Fermi arc states that appear in the surface
spectrum\cite{Arcs}. The Fermi arcs can contribute to the electrical
conductivity in setups where the dominance of the surface transport is
expected, such as thin films and nanowires \cite{TaAs,Cd3As2,NbAsNature}. In
Dirac semimetals, these Weyl points are merged together, and generally
speaking lack topological protection although perturbations that gap them
appear very rarely in real systems\cite{PNAS}. 
\begin{figure}[tbp]
\includegraphics[height=0.58\textwidth,width=0.4\textwidth]{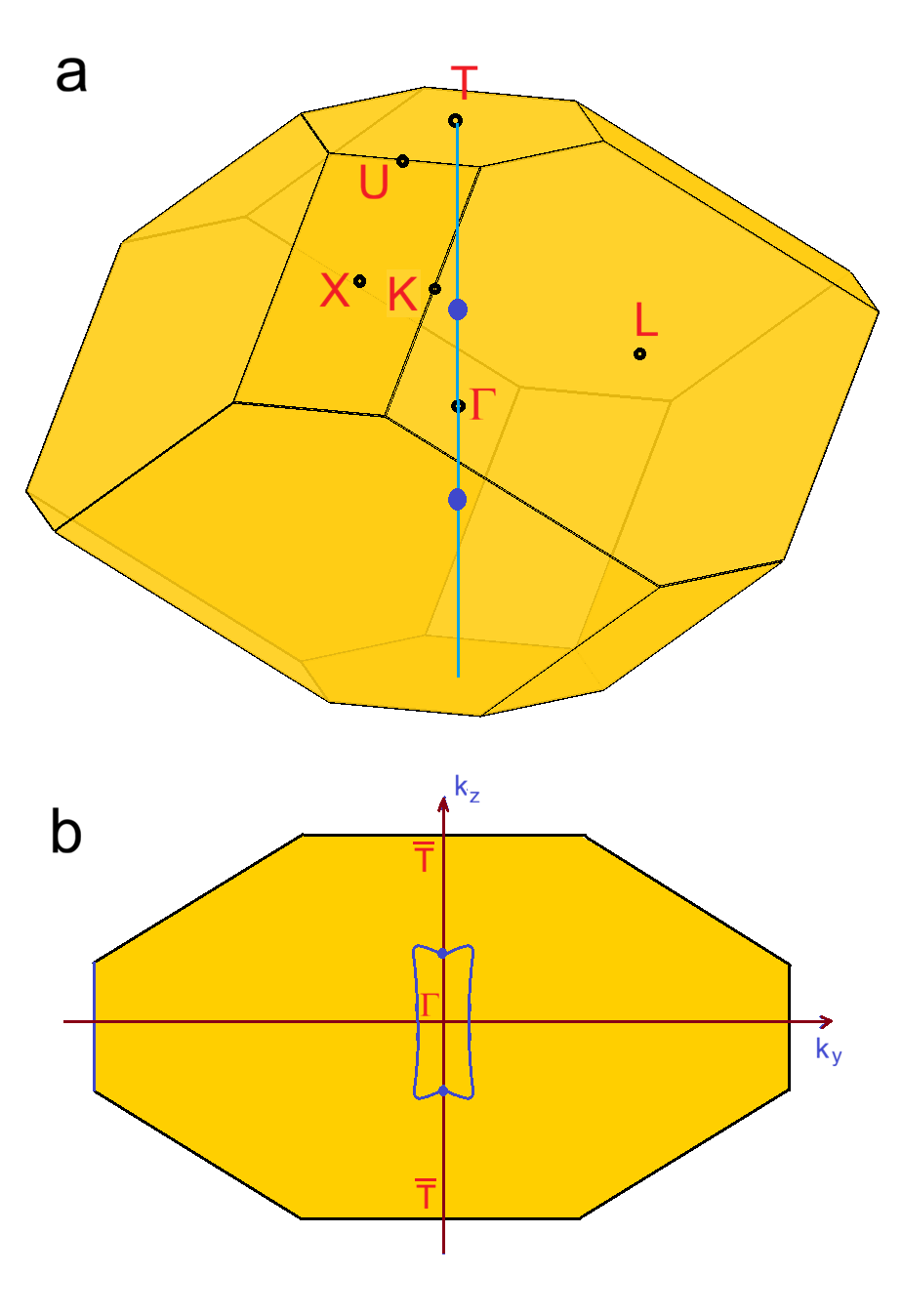}
\caption{a) Brillouin Zone of the rhombohedral lattice of $\protect\beta -$Cu%
$_{2}$Se with high-symmetry points indicated. The Dirac points occur along $%
k_{z}$ direction and are indicated by blue dots. b) Calculated Fermi arc
surface states for $\protect\beta -$Cu$_{2}$Se that are shown in the surface
Brillouin Zone corresponding to the (100) surface.}
\label{FigFermiArcs}
\end{figure}

To shed light whether Fermi arcs are realized in $\beta -$Cu$_{2}$Se, we
perform the band structure calculation for a slab that is oriented along the
100 direction where we expect the two Dirac points to produce two
projections connected by the arcs in the surface BZ. Since self--consistent
slab calculations requiring hundreds of atoms in the supercell are very time
consuming, we take the tight--binding route and utilize exact
non--orthogonal tight--binding transformation of our bulk LDA Hamiltonian
which is possible for the linear muffin--tin orbital basis set employed here%
\cite{FPLMTO,LMTOTB}. This unitary transformation allows us to reproduce the
energy bands of the bulk $\beta -$Cu$_{2}$Se, \ref{FigBands}(a), without
additional approximations. Then, the tight--binding Hamiltonian is extended
to the slab geometry containing 50 original unit cells of $\beta -$Cu$_{2}$%
Se along the x axis (total 300 atoms per supercell) and is exactly
diagonalized to obtain a well convergent surface spectrum.

Analyzing the Fermi surface for the slab, we find two Fermi arcs that
connect Dirac points projected onto (100) surface as illustrated in Fig. \ref%
{FigFermiArcs}(b). From general grounds, each pair of Weyl points of
opposite chirality and located at opposite $\mathbf{k}$ points in the BZ\
should be connected by the arc. In $\beta -$Cu$_{2}$Se, the Dirac points are
made of the merged Weyl points, therefore each pair does not necessarily
lead to its own Fermi arc. In fact, it is known that there are Dirac systems
without arcs \cite{PNAS2}. Nevertheless, two Fermi arcs appear in the
surface spectrum, as seen in Fig. \ref{FigFermiArcs}(b), which means that
the Weyl points located at opposite k points are the sources of arcs.

The Fermi arcs bear the spin texture similar to the helical structure of the
surface spin states in topological insulators: the spins along each arc
point perpendicular to their electron velocities. This pattern changes only
when they approach the surface projections of the Dirac points where local
\textquotedblleft all--in/all--out\textquotedblright\ alignment emerges.

We finally make a note on the surface--mediated transport resulted from the
Fermi arcs. The transport relaxation time for the electrons is usually
described by the scattering on impurities or phonons. Regardless the precise
nature of the scattering matrix elements, we can qualitatively discuss
various scattering contributions by looking into the available phase space.
Let us assume the effects of the scattering inside the bulk and from the
surface to the bulk can be neglected by utilizing thin film geometry. This
is important, since in the thermodynamic limit , the bulk will always have a
nonzero contribution to transport due to thermally excited electrons, and
this will hide all surface phenomena. The main contributions to scattering
are known to be backscattering processes, since for the electronic states at 
$\mathbf{k}$ and $-\mathbf{k}$, their electronic velocities are oppositely
directed. They usually dominate in every three--dimensional Fermi surface
but would disappear in Weyl semimetals since the wave function overlap
between the Fermi arc states at $\mathbf{k}$ and $-\mathbf{k}$ residing on
different surfaces in real space is zero. This, however, is not true for
Dirac semimetals since both Fermi arcs appear at the same surface. We
notice, however, that the spinor states of opposite Fermi arcs are
antialigned due to their particular spin texture. Therefore we expect that
the backscattering effects will cancel each other since the spinors with
oppositely directed spins are orthogonal. Thus, the Fermi--arc surface
states should be contributing to transport in a thin--film--like setup,
which is expected to be strongly anisotropic, have a high carrier mobility
and be resilient to the defects at the surface due to the remarkable
protection of the topological states.

\section{\textbf{Conclusion.}}

In conclusion, based on our density functional calculation, the rhombohedral 
$\beta -$phase of Cu$_{2}$Se is predicted to be an ideal Dirac semimetal
with the Fermi arc states emergent in its surface spectrum. We argued that
the strong suppression of scattering matrix elements in the surface
transport may occur due to the particular shape and spin texture of the
Fermi arcs, resulting in ultra--high carrier mobility and strong anisotropy
of the electrical conductance at the surface.

\end{document}